\documentclass{IEEEtran}
\usepackage{cite}
\usepackage[english]{babel}
\usepackage[normalem]{ulem}
\usepackage{epstopdf}
\usepackage{times}
\usepackage{mathptm}
\usepackage[pdftex]{graphicx}
\usepackage{subfigure}
\usepackage{algorithmic}
\graphicspath {{./figures}}
\usepackage{balance}
\usepackage{refstyle}
\usepackage[perpage,symbol]{footmisc}
\usepackage{graphicx}
\usepackage{epstopdf}
\usepackage{graphics}
\usepackage{amsmath}
\usepackage{epsfig}
\usepackage{balance}
\usepackage[]{algorithm2e}
\usepackage{algorithmic}
\usepackage{color}
\usepackage{multirow}
\usepackage{cite}

\ifCLASSINFOpdf
\else
\fi

\begin{document}
\begin{titlepage}
\title{Secure and Reliable Biometric Access Control for Resource-Constrained Systems and IoT}
\centering
\end{titlepage}

\author{Nima Karimian, Zimu Guo, Fatemeh Tehranipoor, Damon Woodard, Mark Tehranipoor, Domenic Forte
\thanks{Nima Karimian is with the department of Electrical and Computer Engineering, University of Connecticut, USA (e-mail: nima@engr.uconn.edu).}
\thanks{Fatemeh Tehranipoor is with the School of Engineering, San Francisco State University, CA, USA (e-mail: tehranipoor@sfsu.edu).}
\thanks{Zimu Guo, Damon Woodard, Mark Tehranipoor, Domenic Forte are with the department of Electrical and Computer Engineering, University of Florida, USA.}}

\maketitle

\begin{abstract}
With the emergence of the Internet-of-Things (IoT), there is a growing need for access control and data protection on low-power, pervasive devices. Biometric-based authentication is promising for IoT due to its convenient nature and lower susceptibility to attacks. However, the costs associated with biometric processing and template protection are nontrivial for smart cards, key fobs, and so forth. In this paper, we discuss the security, cost, and utility of biometric systems and develop two major frameworks for improving them. First, we introduce a new framework for implementing biometric systems based on physical unclonable functions (PUFs) and hardware obfuscation that, unlike traditional software approaches, does not require non-volatile storage of a biometric template/key. Aside from reducing the risk of compromising the biometric, the nature of obfuscation also provides protection against access control circumvention via malware and fault injection. The PUF provides non-invertibility and non-linkability. Second, a major requirement of the proposed PUF/obfuscation approach is that a reliable (robust) key be generated from the user’s input biometric. We propose a noise-aware biometric quantization framework capable of generating unique, reliable keys with reduced enrollment time and denoising costs. Finally, we conduct several case studies. In the first, the proposed noise-aware approach is compared to our previous approach for multiple biometric modalities, including popular ones (fingerprint and iris) and emerging cardiovascular ones (ECG and PPG). The results show that ECG provides the best tradeoff between reliability, key length, entropy, and cost. In the second and third case studies, we demonstrate how reliability, denoising costs, and enrollment times can be simultaneously improved by modeling subject intra-variations for ECG.
\end{abstract}

%\begin{keywords}

% \fi
\IEEEpeerreviewmaketitle

\section{{Introduction}}
Advances in semiconductors, computing, and networking as well as the upcoming Internet of Things (IoT) have given rise to a ``connected world'' where machine-to-machine, human-to-machine, human-to-thing, etc. interactions are commonplace. Although this is touted to increase convenience and efficiency of everyday life (e.g., through smart devices, vehicles, homes, buildings, cities, healthcare, and more), there is overwhelming evidence that current systems are ill-prepared from a security perspective. This is especially true for low-cost electronic devices or``thing''. For instance, inexpensive hacks have been successfully demonstrated on more than 75\% of Bluetooth low energy (BLE) smart locks in the market~\cite{rose2016picking, Backdooring}. Similarly, IoT wearables and mobile devices have the potential to dramatically lower healthcare costs through remote health monitoring but come with risks of fraud - due to the lack of physical presence at the time of collection, the identity of the user that transmits the health information is indeterminate~\cite{agrafioti2012secure}. Many of the aforementioned problems are related to access control, and can be at least partially alleviated by incorporating biometric authentication. However, doing so successfully can be challenging for resource-constrained IoT devices and applications.

Biometric authentication systems generally consist of five major components: sensor, feature extraction, template storage/database, matcher, and decision module~\cite{jain2008biometric}. The sensor represents the interface between the user and the authentication system, and its function is to scan the biometric trait of the user. Feature extraction processes the scanned biometric data to extract the information that is useful in distinguishing between different users. The feature set extracted during an enrollment phase is either stored in a remote database as a template indexed by the user's identity information (i.e., match-on server) or stored on a smart card (i.e., match-on card/device systems). The matcher may be hardware or software which compares the template with an input query. Finally, the decision module provides a response to the query, i.e., whether the user’s biometric matches the template or not.

Biometric system failures can be categorized into two classes~\cite{jain2008biometric}: (1) intrinsic failures that occur due to intra-user variation and/or lack of distinctiveness for a biometric modality; and (2) failures from attackers who circumvent the system. In the case of the former, processing and matching capabilities of resource-constrained devices may limit decision accuracy of the biometric authentication system. Since IoT systems are supposed to be convenient, users will likely not tolerate high false accept rates which prevent them from accessing their devices and data. The latter (attacks) are an even greater concern. First, many IoT devices may exist in unprotected or hostile environments which leaves the aforementioned components exposed to non-invasive, semi-invasive, and invasive physical attacks that bypass matching or decision modules (e.g., turn a negative decision into a positive one), alter a stored template, perform hill-climbing~\cite{adler2005vulnerabilities}, etc. If successful, an attacker gains illegitimate system access. Second, protection/privacy of the template is an enormous challenge. If the raw biometric template can ever be recovered, it is forever compromised because biometrics are irrevocable. 

Existing countermeasures for protecting the template are limited. In the most common instantiation (match-on server), the template exists on a centralized server in a raw, encrypted, salted, or transformed form. The template or keys/parameters protecting them are vulnerable to hackers who time and again break through security (e.g., Equifax data breach~\cite{Equifax} , OPM theft of 5.1 million fingerprints~\cite{washingtonpost} , etc.). More often than not, communication between devices and servers in IoT applications is insecure (e.g., passwords transmitted in plaintext~\cite{techcrunch}) due to time-to-market constraints or inexperience in security implementations. In an alternative instantiation (match-on card/device), the biometric information never leaves the card/device. However, the aforementioned physical attacks can be used to gleam the template from the card/device or bypass critical security modules. Recently, homomorphic encryption has been proposed for securing biometric templates because it performs matching in the encrypted domain~\cite{cheon2016ghostshell,karabat2015thrive}. However, such approaches are still far from efficient, requiring several orders of magnitude extra hardware and operational time~\cite{Cio}. Further, they still may be susceptible to attacks on the secret key kept on the user-side and fault injection at the decision module.

In this paper, we propose a new paradigm called ``\underline{B}iometric \underline{L}ocking by \underline{O}bfuscation, Physically Un\underline{c}lonable \underline{Ke}ys, and \underline{R}econfigurability'' (BLOcKeR) which aims to providie low-cost template protection and attack resistance in match-on card/device applications. A central element of BLOcKeR is hardware personalization through reconfigurability. In short, the cost associated with biometric systems is improved by adapting the preprocessing, feature extraction, and postprocessing modules on a user-to-user basis. In addition, we combine two advances in hardware security – obfuscation~\cite{forte2017hardware} and physical unclonable functions (PUFs)~\cite{gassend2002silicon, suh2007physical} – with configurability to significantly increase resistance to various attacks against biometric systems. Our main contributions can be summarized as follows: 
\begin{enumerate}
\item \textbf{BLOcKeR}: We discuss the high-level features of BLOcKeR, compare its features to traditional schemes, and analyze its security against various attacks. Hardware obfuscation removes the matcher and decision modules, thus substantially reducing vulnerabilities. A PUF, which is essentially a hardware biometric, is added in order to avoid permanently storing the human biometric and/or cryptographic keys on the device. The PUF also provides a non-invertible transform to the biometric in order to protect the biometric template and make the biometric/obfuscation non-linkable to different devices. 
\item \textbf{Noise-aware Feature Quantization}: We propose NA-IOMBA to select and quantize biometric features on a user-to-user basis. In other words, only the most robust features are selected for each user, thus avoiding unnecessary post-processing costs. When available, noise models are used to predict the impact of noise and further reduce error correction costs and enrollment time.
\item \textbf{Case Studies}: We perform three case studies to demonstrate the effectiveness of NA-IOMBA. Experimental results show that NA-IOMBA increases the length, entropy, and robustness of keys generated from multiple biometric modalities (fingerprint, iris, ECG, PPG). ECG noise models for baseline wander, motion artifacts, EMG noise, and stress/exercise are adopted to improve ECG key reliability and reduce error correction costs. Implementation results from FPGA show that, even with less resources (65\% power and 62\% utilization),  reliability is improved after integrating noise models into NA-IOMBA. 
\end{enumerate}

The rest of the paper is organized as follows. We introduce hardware obfuscation, PUFs, the BLOcKeR flow, and  its merits in Section 2. In Section 3, the noise-aware quantization framework is described. NA-IOMBA case studies including comparison multiple biometric modalities, reduction in denoising overhead, and reduction in enrollment times are discussed in Section 4. In Section 5, conclusions are provided.

\section{Hardware Security Preliminaries and BLOcKeR}
\label{sec:BLOcKeR}

\label{sec:hardware primitives}
\subsection{Hardware Obfuscation}
\label{sec:obfuscation}
Hardware obfuscation encapsulates a series of techniques which lock a chip or system by blocking its normal function at the hardware level until a correct key is applied. Without a correct key, the device can be referred to as `locked’ or `obfuscated’. In the typical instantiation, only the designer can compute the key that unlocks the device. Obfuscation techniques can be classified into three categories: logic encryption (locking), logic permutation, and finite state machine (FSM) locking.

\begin{figure}[t]
\centering
\includegraphics[width=\linewidth]{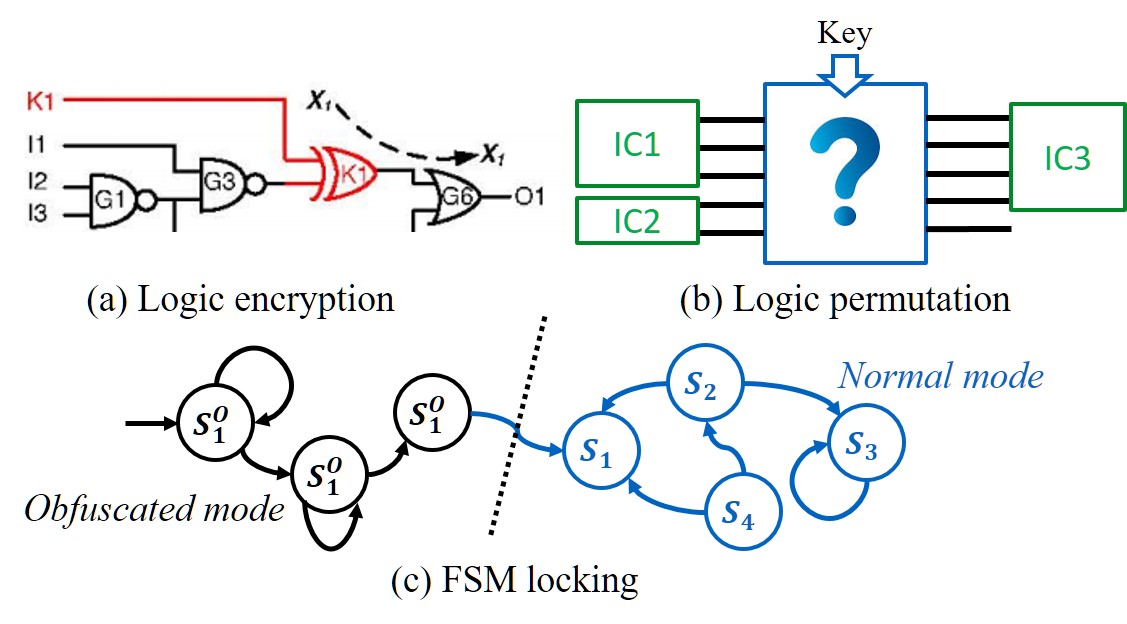}
\small
\caption{(a) Logic encryption~\cite{rajendran2012logic}; (b) logic permutatio, (c) FSM based on obfuscation.}
\label{fig:obfuscation}
\end{figure}

The logic encryption approach operates by inserting additional components into the internal paths of combinational logic in the original (un-obfuscated) design. These paths indicate the gate-level interconnections as shown in Fig. \ref{fig:obfuscation}(a). The additional components, such as exclusive or (XOR) and exclusive nor (XNOR) gates, modify the internal value of a path based on the key. If the key is incorrect, the logic value becomes inverted, thus creating an incorrect behavior in the design. Typically, a large number $n$ of additional components (32, 64, 128, etc.) are inserted; thus, creating $2^n$ possible key combinations. Besides XOR/XNOR gates, modified look-up table (LUTs), which can involve even more key bits per additional component, can be utilized. Compared with logic encryption, logic permutation alters the order of the interconnections instead of changing their values. An example of system-level obfuscation is presented in Fig. \ref{fig:obfuscation}(b). The key-controlled multiplexer (question mark box) conceals the interconnections among integrated circuits (ICs). The finite-state machine (FSM) based technique \cite{chakraborty2009harpoon} embeds additional states into the original FSM as shown in Fig. \ref{fig:obfuscation}(c). In order to unlock the design, a designated key input pattern navigates the FSM from the obfuscated state space to the original state space where the device is unlocked.

\begin{figure}[t]
\centering
\includegraphics[width=0.6\linewidth]{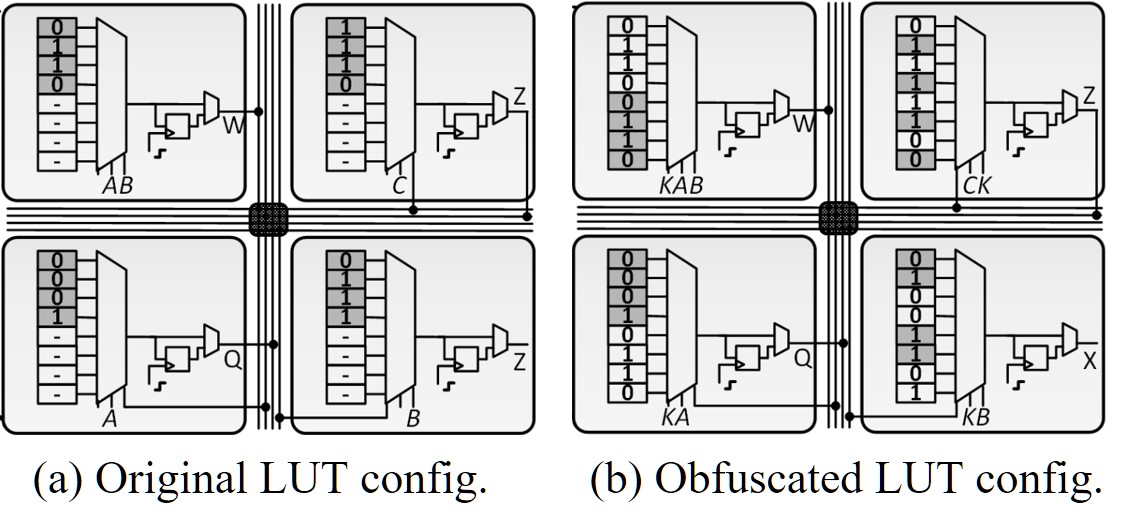}
\small
\caption{Robust bitstream protection in FPGA-based systems through low-overhead obfuscation \cite{karam2016robust}.}
\label{fig:lut obfu}
\end{figure}

In BLOcKeR, we utilize the bitstream obfuscation scheme which was first introduced in \cite{karam2016robust}. This key-based bitstream obfuscation technique applies to reconfigurable fabrics and can be used with off-the-shelf field programmable gate array (FPGA) hardware. It leverages unused FPGA resources within an existing design as shown in Fig. \ref{fig:lut obfu}, thereby incurring very little overhead. In this figure, the original design consists of 4 configurable logic blocks (CLBs) which can support three input functions. However, only two input functions are used by the original application, thus leaving half of the LUT unoccupied. After obfuscation, these unused bits are filled with other functions; the correct response selection depends on the key input position and value. The obfuscated bitstream can take one of two formats: (a) structural Verilog, which implements the circuit as a series of assignment statements, or (b) using device-specific LUT primitive functions. Compared to the previously described obfuscation approaches, we choose bitstream obfuscation because it allows the obfuscation key to change for every chip/user due to the reconfigurable hardware. This is an important feature since a biometric-derived key will be different from user to user. Other obfuscation approaches are fixed by a master key and cannot be adapted in this manner.

\subsection{Physical Unclonable Functions (PUFs)}
\label{sec:PUF}
The best way to describe a PUF is that it is ``an object's fingerprint''. Like a biometric which expresses a human's identity, a PUF is an expression of an inherent and unclonable instance-specific feature of a physical object \cite{maes2016physically}. For integrated circuits (ICs), these instance-specific features are induced by manufacturing process variations and can be captured by the input/output or challenge/response pairs (CRPs) of the PUF. The challenge/response relationship can be viewed as a physical one-way function which provides the design with a non-invertible capability. In addition, due to IC-to-IC process variations, the CRPs are unique to each device. We refer to this property as non-linkability.

A variety of PUFs have been proposed in the literature \cite{maes2016physically}. They can be broadly classified into electronic and non-electronic categories. In the former, the challenge-response mechanism is determined based on the electronic properties of an object such as delay of internal circuit paths, the threshold voltages of transistors, etc. Examples of this PUF type include arbiter PUF, ring oscillator PUF, SRAM PUF, etc. The non-electronic PUF generates the CRPs based on the non-electronic properties of an object, such as magnetic, acoustic, radio frequency, etc.

According to the number of total CRPs available, PUFs can be classified as \textbf{strong} and \textbf{weak}. A PUF is called strong if, even after giving an adversary access to a PUF instance for a prolonged period, it is still possible to come up with an unknown CRP with high probability. This implies that the considered PUF should have an exponentially sized CRP set. Otherwise, the adversary can simply query all challenges. PUFs which do not meet these requirements are consequentially called weak PUFs. In BLOcKeR, a strong PUF is utilized.

A recently developed attack against the strong PUF applies well-developed machine learning algorithms (such as logistic regression) to model challenge-response behaviors \cite{ruhrmair2010modeling} based on a subset of CRPs. This CRP set is collected from the strong PUF via the direct physical access. With this model, a new CRP can be predicted without having physical access to the device. A high prediction rate (i.e., 99.9\%) can be achieved with short training time (i.e., 2.1 seconds) \cite{ruhrmair2010modeling}. \textit{Note that, in BLOcKeR, this modeling attack is seen as a benefit. More details will be given in Section \ref{sec:authentication flow}.}

\subsection{BLOcKeR Flow}
\label{sec:authentication flow}
In this section, the BLOcKeR enrollment and authentication processes are discussed (see Fig. \ref{fig:BLOcKeR_flow}). The enrollment process involves three major steps: hardware enrollment, ownership claim, and firmware customization.

The first step to initialize BLOcKeR, \textbf{hardware enrollment}, is accomplished before the device is sent to the market and occurs in a trusted environment. During this step, the designer or system vendor builds a strong PUF model for each device using a dedicated firmware and stores the models in a secure database. This firmware enables the designer to efficiently collect a sufficient number of CRPs for the prediction model. After enrolling the PUF model, this firmware will be removed. At this point, the device encloses no firmware and will be sold to the user through the insecure supply chain. Since neither the user nor attacker could have the high-speed and direct access to the PUF challenges after this point, the prediction model is therefore accessible only to the designer.

To register and operate the device, the \textbf{ownership claim} step is taken by the legitimate user\footnote{Note that this step only occurs once}. The user’s ownership is taken by presenting his/her biometric signal to the device. A pre-processing algorithm is applied on the received biometric to extract the binary \textit{bio\_key}. Along with this process, necessary helper data (i.e., error correction code or ECC \cite{karimian2017highly}) might be generated for correcting errors during later authentication steps. A PUF is used as a one-way transform on the quantized biometric. In order to generate the PUF challenge, the \textit{bio\_key} is processed by a hardware hash function to the desired length. The output is transmitted to the designer through a secure channel. This can be achieved by either a trusted retailer (e.g., Verizon) or a trusted platform module (TPM) enabled device (e.g., laptops). \textit{Note that in BLOcKeR, the biometric template is never stored on the device and never sent to the designer/vendor.}

When the PUF challenge is received by the designer, the \textbf{firmware customization} step occurs. This is where the previous strong PUF model is beneficial. The challenge is fed into the strong PUF model to compute a unique device and biometric dependent response, which will behave as an obfuscation key (\textit{obs\_key}). An obfuscated bitstream is produced that will exploit this obfuscation key. The obfuscated bitstream is sent to the user and loaded into the device. Note that since the physical device with the PUF is no longer accessible to the designer/vendor, this step would not be possible without the previously generated strong PUF model.

During the authentication process, the user provides his/her biometric as input. The same pre-processing algorithm as the enrollment process is applied to generate the \textit{bio\_key}. Potential errors are corrected with the helper data. Next, the hash function creates the challenge. The \textit{obs\_key} is then generated by injecting this challenge into the strong PUF. Different from the enrollment process, this obfuscation key is generated on the physical device instead of its mathematic model. A correct \textit{obs\_key} unlocks the obfuscated bitstream and brings the device into functional (unlocked) mode. Without the correct key, the device will simply not work correctly. For example, it will be unable to access data, perform critical protocols, etc. Note that the biometric template, \textit{bio\_key}, and \textit{obs\_key} are never stored in non-volatile (permanent) memory on the card/device or server.

It’s important to note that since the obfuscation key is generated from the PUF circuit, it may be subjected to various environmental noise such as temperature instability, supply voltage fluctuation, etc. For instance, the arbiter PUF exploits a race condition between two paths of the same length. Process variations of the transistors on these paths determine the competition result. However, an increased temperature could cause a faster path to become slower. In some instances, the temperature effect overrides the process variation and generates errors. Thus, the PUF response may vary slightly even for the same challenge due to noise. To address these errors, an additional ECC module can be implemented for the PUF. Alternatively, the obfuscated bitstream can be constructed to tolerate these errors. The latter is a promising topic for future work.

\subsection{Salient Features of BLOcKeR}
\label{sec:beyond attacks}
The following are noteworthy features of the BLOcKeR framework:
\begin{itemize}
\item By incorporating a PUF, the scheme possesses a \textbf{non-invertible property} that makes it similar to feature transformation approaches in the literature~\cite{jain2008biometric}. This implies that an attacker cannot reconstruct the biometric template even if the obfuscation key is ever stolen from the server or working card/device.
\item Another major advantage of the PUF is its \textbf{non-linkable property}, which means that a user's biometric can be only used to unlock one device. In other words, the obfuscated bitstream produced for every device is unique to both the device and the legitimate user. Therefore, even a legitimately enrolled user of one device cannot unlock the device which it is not bonded with.
\item The enrollment/authentication processes inherently provide \textbf{information security}. All private information is kept secret and never stored in a non-volatile (permanent) memory. Only the strong PUF challenge is transmitted to the server and the biometric template is protected by non-invertibility of the hash function and PUF. The obfuscation key is generated when the user authenticates and is also never permanently stored. These features dramatically reduce the attack surface at the chip/device, server, and communication channel (more details in next section).
\item \textbf{Reconfigurability } is advantageous because it allows each device’s obfuscation/locking to be tuned to its owner. In addition, it allows the hardware to adapt pre-processing, feature extraction, post-processing, and error correction overheads on a user-to-user basis. This latter benefit will be made clearer in the remaining sections of the paper.

\end{itemize}

\subsection{Attacks Analysis}
\label{sec:attacking analysis}
As discussed earlier, current biometric technologies can be classified into match-on card/device and match-on server. Considering the cost and accessibility of the device under attack, the attacks can be labeled as follows. \textit{Non-invasive attacks}~\cite{weber2005software, yang2004scan} require the lowest cost and no physical tampering. \textit{Semi-invasive attacks}~\cite{skorobogatov2005semi, skorobogatov2010flash} require intermediate cost and minimal physical tampering (e.g., backside thinning). \textit{Invasive attacks}~\cite{marinissen2014direct} require the highest cost and full physical tampering.

In Table \ref{tab:attacks}, the attacks are organized by their cost and applicability. The red italic text indicates the attacker's objective – template theft/recovery, denial of service, or unauthorized card/device access. The invasive attacks such as microprobing~\cite{marinissen2014direct} can be utilized to compromise the non-volatile memory, such as Flash or Anti-fuse, on the card/device. Semi-invasive attacks, such as the fault injection and memory bumping attack~\cite{skorobogatov2005semi}, can be applied to create false accepts by the matcher and decision module or to determine the template respectively. Other semi-invasive attacks, such as modifying templates, may cause the denial of service or unauthorized access. Since both invasive and semi-invasive attacks require at least minimal physical access, these attacks can be only applied on the match-on card/device frameworks. In contrast, non-invasive attacks apply to both types of frameworks. Software-level attacks~\cite{weber2005software}, such as software Trojan, can be utilized to leak biometric templates. The hill-climbing attack is performed by iteratively submitting synthetic representations of the user's biometric until a successful recognition is achieved~\cite{maiorana2015hill}. At each step, the employed data are modified according to the matching score of previous attempts. If the cryptography techniques are engaged to protect the template, then side-channel attacks may be employed. Side channel attacks rely on the information gained from data and control flow dependent emanations from the cryptosystem's physical implementation~\cite{kelsey1998side}, such as power, timing, electromagnetic, acoustic, etc. Finally, the communication channel from/to the server is also vulnerable when the protocol is not properly implemented.

The unique structure of BLOcKeR protects it from all the above attacks. Since the template is never stored anywhere, the template privacy issues are eliminated. In BLOcKeR, no matcher or decision module is implemented. Instead of comparing the key with a template, an obfuscation key resolves a device-unique obfuscation. Thus, no semi-invasive attacks, which target these components, are applicable.
\begin{figure}[h]
\centering
\includegraphics[width=0.7\linewidth]{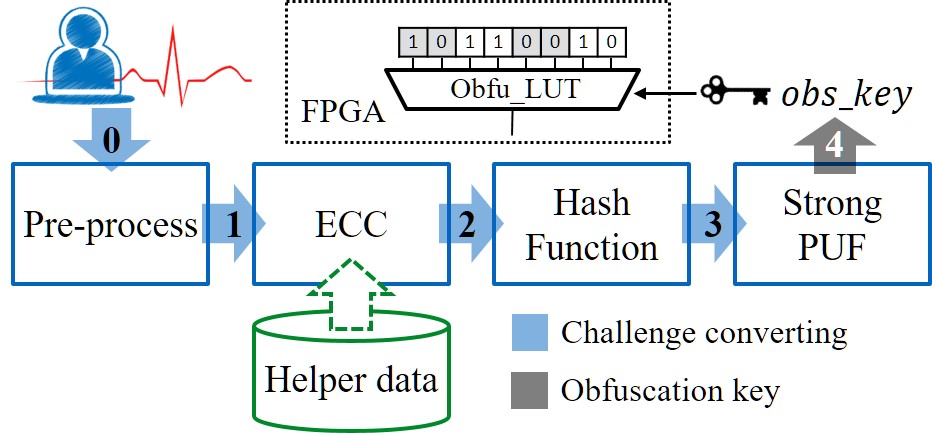}
\caption{Overall implementation flow of BLOcKeR.}
\label{fig:BLOcKeR_flow}
\end{figure}
Since the bitstream obfuscation is engaged in BLOcKeR, obfuscation-oriented attacks must also be accounted for since they may introduce new vulnerabilities. The most straightforward attack is brute-force. An adversary either randomly applies: (i) synthesized biometric signals (at arrow 0 in Fig. \ref{fig:BLOcKeR_flow}) or (ii) the obfuscation key (at arrow 4 in Fig. \ref{fig:BLOcKeR_flow}) until the correct system functionality is observed. This attack is infeasible in both cases due to the extremely large analog input and binary obfuscation key spaces. Furthermore, brute force cannot be improved through hill climbing since (i) there is no matching module to provide a score and (ii) a non-invertible transform is applied to the input. Specifically, the data flow back from arrow 4 to 3 (Fig. \ref{fig:BLOcKeR_flow}) is blocked by the strong PUF. The data flow back from arrow 3 to 2 (Fig. \ref{fig:BLOcKeR_flow}) is blocked by the one-way hash function.

Besides brute force, a more advanced attack on obfuscation is based on satisfiability checking (SAT) \cite{subramanyan2015evaluating}. The SAT attack infers the correct key using a small number of carefully selected input patterns and their correct outputs observed from an activated functional device. Since the obfuscation key is destroyed after losing power (i.e., it's not stored in non-volatile memory), it is unlikely for an adversary to capture an activated device. Even if the adversary is a legitimate user and has access to an activated device, the strong PUF module is still a black box. Thus, it is impractical for the adversary to generate a desired obfuscation key for another device using synthetic biometric signals.

\begin{table}[h]
\centering
\scriptsize
\renewcommand\arraystretch{1.3}
\caption{Possible attacks and vulnerabilities of current employed biometric authentication strategies.}
\label{tab:attacks}
\begin{tabular}{|c|c|c|}
\hline
& Match-on server & Match-on card \\ \hline
\begin{tabular}[c]{@{}c@{}}Invasive \\ attacks\end{tabular} & -- & \begin{tabular}[c]{@{}c@{}}Micro-probing,\\ \textcolor{red}{\textit{Template privacy}}\end{tabular} \\ \hline
\multirow{2}{*}{\begin{tabular}[c]{@{}c@{}}Semi-invasive\\ attacks\end{tabular}} & \multirow{2}{*}{--} & \begin{tabular}[c]{@{}c@{}}Fault injection;\\ Memory bumping\\ \textcolor{red}{\textit{Unauthorized access}}\end{tabular} \\ \cline{3-3}
& & \begin{tabular}[c]{@{}c@{}}Modify template/parameter\\ \textcolor{red}{\textit{Denial of Service}} \\ \textcolor{red}{\textit{\& Unauthorized access}}\end{tabular} \\ \hline
\begin{tabular}[c]{@{}c@{}}Non-invasive\\ attacks\end{tabular} & \begin{tabular}[c]{@{}c@{}}Buffer overflow;\\ Software Trojan;\\ Malware;\\ Hill climbing;\\ Comm. protocol\\ \textcolor{red}{\textit{Template privacy}}\end{tabular} & \begin{tabular}[c]{@{}c@{}}Buffer overflow;\\ Software Trojan;\\ Malware;\\ Hill climbing;\\ Side-channel attack\\ \textcolor{red}{\textit{Template privacy}}\end{tabular} \\ \hline
\end{tabular}
\end{table}

\section{Noise-Aware Quantization Framework}
\subsection{IOMBA based Biometric Key Generation}
Biometric systems that operate using any single biometric characteristic have the several limitations such as noise in sensed data, intra-class variations (the biometric data acquired from an individual during authentication may be very different from the data that was used to generate the template during enrollment), and distinctiveness (biometric trait is expected to vary significantly across individuals). Noise in biometric authentication and recognition applications can be tolerated to some extent; however, not a single error can be tolerated in key generation applications like BLOcKeR. To address this limitation, we have previously developed the interval optimized mapping bit allocation (IOMBA) scheme for biometric key generation ~\cite{karimian2017highly}.

In short, IOMBA tunes the biometric key generation process to each user rather than relying on a generic approach for all users. The performance is controlled by two parameters, $\alpha$ and $\beta$, which define the entropy and reliability of generated keys respectively. IOMBA eliminates features from the space that have low entropy and quantizes features with larger entropy and lesser noise into more bits. Then, since noise will impact each user’s biometric differently, the features that are more sensitive to noise (i.e., unreliable for key generation) are also removed on a user-to-user basis. As a result, the length of keys generated varies based on the $\alpha$ and $\beta$ parameters as well as from user to user.

The major steps of the IOMBA are described as follows. 

\textbf{1) Data Pre-Processing:} The signals from the population are pre-processed to remove noise followed by feature extraction. The feature elements from the same location are extracted from the population and normalized into a standard normal distribution. The same normalization parameters are later exploited to normalize the corresponding feature elements of each subject. Our approach can work with any biometric provided it produces statistically independent and Gaussian features in some representation. Note that any feature that does not meet these requirements is discarded.

\begin{figure}[h!] 
\centering
\small
\includegraphics[width=\linewidth]{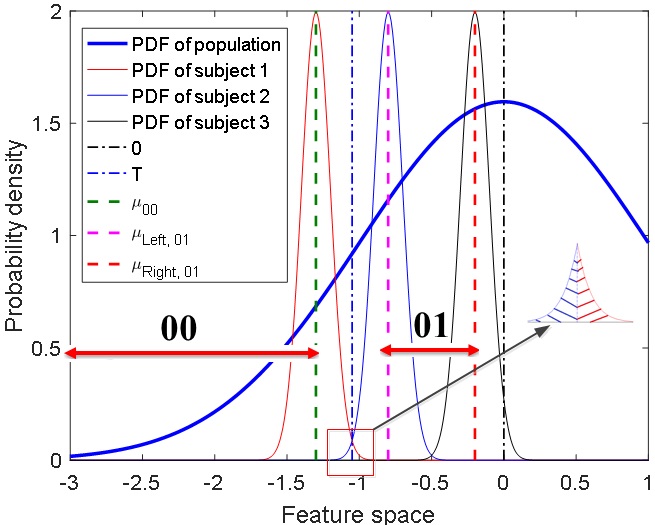}
\caption{Schematic for optimization of quantization with two bits.}
\label{pdf}
\end{figure}

\textbf{2) IOMBA Margin Calculation from Population Statistics:} IOMBA quantizes each feature into a different number of bits. $2$ bit quantization is illustrated in Fig.~\ref{pdf}. The population $PDF$ of a feature is shown in blue. As an exampleexample, to illustrate the reliability calculation, three subject’s distributions are considered. According to these distributions as well as the $\alpha$ (entropy) and $\beta$ (reliability) parameters, the boundaries ($\mu_{left,01}$, $\mu_{right,01}$, $\mu_{00}$) and threshold ($0$, $T$) which satisfy the following equations are computed 

{ \small
\begin{equation}
\small
\frac{\int_{\infty}^{\mu_{00}}PDF_{pop}(x)dx}{\int_{\mu_{left,01}}^{\mu_{right,01}}PDF_{pop}(x)dx} \leq \alpha \\
 \int_{T}^{\infty}PDF_{sub_1}(x)dx \leq \beta 
\end{equation}}

{\small
\begin{equation}
\small
\label{eq:beta}
 \\
\int_{-\infty}^{T}PDF_{sub_2}(x)dx \leq \beta \\ 
\int_{0}^{\infty}PDF_{sub_3}(x)dx \leq \beta
\end{equation}} 
$ PDF_{pop}(x)$ denotes the distribution for feature $x$ for the entire population and $ PDF_{sub_y}(x)$ denotes the distribution (due to noise) for feature $x$ of subject $y$. Fig.~\ref{pdf} illustrates these values on the negative side of the $x$ axis. The positive boundaries and thresholds can be simply computed by mirroring the negative values onto the positive side against the $y$ axis. These boundaries and thresholds guarantee that enrolled and regenerated key bits are statistically random and reliable. 

\textbf{3) Enrollment for Key and Helper Data Generation:} Essentially, IOMBA personalizes a biometric system to each user. For each subject, the key generation framework utilizes the above boundaries to determine whether each feature element is good for generating key bits or not. The least reliable features of the user which do not fulfill the above constraints are discarded. The helper data for each subject consist of the following: (i) the index of the reliable features selected for the user, (ii) the number of bits each feature can be quantized into, and (iii) the normalization parameters for each feature. If error still exists then error correction based helper data and an error correction module can also be added.

\textbf{4) Key Regeneration:} The user presents his/her biometric, features are extracted, and the helper data stored on the system is used to eliminate the unreliable features and then quantize the reliable features to regenerate the key. 

In IOMBA, feature extraction, helper data, and key regeneration steps are different for each user. Reconfigurable hardware is therefore a natural candidate for implementing IOMBA because each step can be tailored uniquely to the user, thereby avoiding certain processing costs.

\subsection{Noise Aware IOMBA (NA-IOMBA)} In the original IOMBA, standard deviation in user PDFs was fixed for each feature in a worst case manner. Feature selection was therefore pessimistic and resulted in shorter keys. Estimating the standard deviation, especially for continuous biometrics (keystroke dynamics, ECG, etc.), is nontrivial since the biometric would need to be collected at all conditions and types of noise. In most applications, such as IoT, the enrollment process would be too long for users to tolerate. In addition, the impact of noise is substantially effected by the type and amount of pre-processing. In resource constrained scenarios, it would be better to eliminate pre-processing steps which are costly and energy consuming to perform. To accommodate these issues, we propose the noise-aware IOMBA framework in this section and demonstrate its benefits in Section 4. NA-IOMBA is a variant of IOMBA that incorporates models to predict the impact of different noise sources, noise scale, and pre-processing steps on biometric key generation. In short, all IOMBA steps are performed with one major change. Margins and boundaries are recomputed based on more accurate estimates of user feature standard deviation. Features that are modeled as less (more) susceptible to noise will therefore be given smaller (larger) margins than IOMBA. The following are the impacts and benefits of this approach:
 
\begin{itemize}
\item \textbf{Impact on key length}: If a margin for a feature increases, it could result in the feature \textbf{being selected} and/or longer bit lengths compared to IOMBA. If a margin for a feature shrinks, it could result in the feature \textbf{no longer being selected} and/or shorter bit lengths compared to IOMBA.

\item \textbf{Impact on key reliability}: If provided noise samples or models are accurate, the reliability of key should improve regardless of whether or not the key length shrinks/grows. 

\item \textbf{Impact on cost and enrollment time}: There are three ways that overheads can be reduced. First, error correction costs tend to increase nonlinearly. By improving key reliability, error correction hardware can be substantially reduced. Second, certain denoising/filtering steps can be removed provided that NA-IOMBA models accurately estimate the noise appearing in features without them. Third, number of enrollment samples and enrollment time can be reduced if noise over time and different conditions can be modeled. This can be particularly important for continuous physiological signals, like ECG and PPG, which can be impacted by so many different conditions, e.g., exercise, stress, and food/drink/drug consumption.
\end{itemize}
\section{NA-IOMBA Case Studies}
The above impacts will be demonstrated in the next section.
\subsection{Case Study I: Comparison of Multiple Modalities Using IOMBA and NA-IOMBA}
In this section, we present a comprehensive performance evaluation of biometric-based key generation. We apply our approaches (IOMBA \& NA-IOMBA) on four biometric modalities: ECG, PPG, iris, and fingerprint. Table~\ref{bio-tech} shows the methodologies, databases, and train/test sizes, that have been employed for multiple biometric modalities.

\begin{table*}[t!]
\centering
\small
\caption{High-level summary analysis for multiple biometric modalities}
\label{bio-tech}
%\scalebox{0.52}{
\begin{tabular}{c|c|c|c|c|c|}
\cline{2-6}
& Databases & Population size & Train / Test sizes & Pre-processing & Feature Extraction \\ \hline \hline
\multicolumn{1}{|c|}{ECG} & PTB & 52 & 52*1000/1560*1000 & FIR filter, R peak det., Segmentation & NCN \\ \cline{1-1}
\multicolumn{1}{|c|}{PPG} & Capnobase & 42 & 42*700/ 840*700 & Butterworth, Peak det., Segmentation & DWT \\ \cline{1-1}
\multicolumn{1}{|c|}{Iris} & CASIAV1 & 108 & 108*4800/756*4800 & Localization & Gabor wavelet\\ \cline{1-1}
\multicolumn{1}{|c|}{Finger} & FVC2004-DB3 & 100 & 100*6056/800*6056 & Normalization, Orientation & Gabor wavelet \\ \hline
\end{tabular}
%}
\end{table*}

\begin{table*}[th]
\centering
\caption{IOMBA/NA-IOMBA Results for Four Biometric Modalities}
\label{tab1}
%\scalebox{0.55}{
\begin{tabular}{clccclccccccc}
\cline{3-13}
& & \multicolumn{3}{c}{Key Length ($N$)} & & \multicolumn{3}{c}{Reliability} & \multicolumn{1}{l}{} & \multicolumn{3}{c}{Min-entropy} \\ \cline{3-5} \cline{7-9} \cline{11-13} 
& & Max & Ave & Min & & Max & Ave & Min & \multicolumn{1}{l}{} & Max & Ave & Min \\ \hline \hline
\multirow{2}{*}{ECG} & NA-IOMBA & 1247 & 953 & 784 & & 100 & 98.76 & 96.17 & & 1.000 & 0.9843 & 0.9055 \\
& IOMBA & 976 & 668 & 512 & & 100 & 97.93 & 94.71 & & 1.000 & 0.9819 & 0.8904 \\ \hline
\multirow{2}{*}{PPG} & NA-IOMBA & 195 & 107 & 17 & & 100 & 99.04 & 91.11 & & 0.9943 & 0.8630 & 0.6229 \\
& IOMBA & 175 & 114 & 14 & & 100 & 96.63 & 89.00 & & 0.9787 & 0.8047 & 0.5732 \\ \hline
\multirow{2}{*}{Iris} & NA-IOMBA & 1136 & 556 & 128 & & 100 & 98.36 & 93.42 & & 1.000 & 0.9376 & 0.7870 \\
& IOMBA & 204 & 66 & 23 & & 100 & 96.81 & 86.73 & & 1.000 & 0.869 & 0.7564 \\ \hline
\multirow{2}{*}{Finger} & NA-IOMBA & 3567 & 1004 & 321 & & 99.54 & 98.76 & 91.72 & & 1.000 & 0.8219 & 0.4913 \\
& IOMBA & 1164 & 835 & 187 & & 99.20 & 95.12 & 72.29 & & 0.9939 & 0.7574 & 0.3258 \\ \hline
\end{tabular}
%}
\end{table*}
\noindent \textbf{Electrocardiogram (ECG)}: 
ECG is a recording of the electric potential, generated by the electric activity of the heart. The ECG recordings of 52 subjects from the PTB database~\cite{goldberger2000physiobank} are used in this paper. We employ low and high pass finite impulse response (FIR) filters with cut off frequencies 1Hz-40Hz to eliminate noise associated with an ECG signal. Normalize-Convoluted Normalize (NCN) is used as the feature extraction technique~\cite{karimian2017noise}.

\noindent \textbf{Photoplethysmogram (PPG)}: 
The photoplethysmogram (PPG) is a biomedical signal that estimates volumetric blood flow changes in peripheral circulation using low-cost and simple LED-based devices typically placed on the fingertips. In order to evaluate the efficiency of the PPG biometric authentication based on IOMBA and NA-IOMBA, a publicly available Capnobase dataset~\cite{karlen2013multiparameter} with 42 subjects was used. We adopt the pre-processing, feature extraction from~\cite{karimian2017human}.

\noindent \textbf{Iris}: 
The iris is the annular region of the eye bounded by the pupil and the sclera (white of the eye) on either side. To evaluate the iris key generation based on IOMBA and NA-IOMBA, we first take the iris images from available CASIAv1-Interval iris database~\cite{CASIA}.~\cite{daugman2004iris} approche is used for pre-processing stage, and fearure extraction.

\noindent \textbf{Fingerprint}: 
A fingerprint is the pattern of ridges and valleys on the surface of a fingertip, the formation of which is determined during the first seven months of fetal development. The fingerprint used for biometric key generation for system authentication in the study is taken from FVC2004 database~\cite{maio2004fvc2004}. In this paper, Gabor filter is used to directly extract fingerprint features from gray level images~\cite{ross2003hybrid}.

The quality of generated keys are compared by four evaluation criteria: reliability, entropy, key length, and cost. The metrics used for each are discussed below and a brief comparison of the above biometric modalities based on IOMBA and NA-IOMBA is provided in Tables~\ref{tab1} under reliability, entropy, and key length. In the table, `max', `ave', and `min' columns correspond to the highest value (best case) achieved among all users, the average of keys across the users, and the lowest value (worst case) achieved among all users.  Note that for this initial comparison, the noise model in NA-IOMBA is adopted from standard deviation of enrollment measurements and standard deviation is adjusted on a per feature basis.  A more elaborate model will be used for ECGs only later in the paper.

\noindent \textbf{Reliability}: Reliability of key generation represents the stability of keys over time. 
If all bits generated by the biometric of an individual are equal to the key produced in enrollment, it can be considered as reliable. We adopt the Reliability metrics from~\cite{karimian2017highly}. As can be seen in Table~\ref{tab1}, improvements in reliability are achieved by applying the NA-IOMBA technique. Average and worst cases improve by 2\% and 9.7\% on average for all modalities compared to IOMBA. Among all modalities, fingerprint attains the largest percentage improvements (3.8\% and 26.9\% in average and worst cases). However, ECG has the best performance for both NA-IOMBA and IOMBA.

\noindent \textbf{Entropy}: To measure key randomness, we calculate the min-entropy.We adopt the Min-entopy metrics from~\cite{karimian2017highly}. As shown in Tables~\ref{tab1}, the min-entropy of ECG signal is higher than iris, PPG and fingerprint However, under NA-IOMBA technique, there is a huge entropy improvement for iris, fingerprint and PPG compare to IOMBA results. For example, the min-entropy is not only improved by 35\% at minimum case for fingerprint based on NA-IOMBA, but also increased by 8\% at average case.

\noindent \textbf{Key length}: Since certain features may be reliable for some users and unreliable for others, our approach will only use reliable features from each individual. Thus, the key length per person may change. As can be indicated in Table~\ref{tab1}, the key length of ECG, PPG, iris, and fingerprint based on IOMBA are 668, 114, 66, and 835, respectively. When NA-IOMBA is applied, the average key length for ECG, PPG, iris, and fingerprint increases by 30\%, 6\%, 88\%, and 27\%. Fingerprint obtains the largest key for both IOMBA and NA-IOMBA while PPG obtains the smallest.

\noindent \textbf{Cost}: The effectiveness of a biometric technology is dependent on how and where it is used. Each biometric modality has its own strengths and weaknesses. Today, an ECG or a PPG sensor costs around $\$20$ when ordered in large quantities, thus has marginal cost of embedding into a biometric system. However, fingerprint and iris scan costs about $\$70$ and $\$ 280$, respectively. Note that the hardware cost is normalized into $1$ in order to make it simpler to consider as a metrics. In that case, if the value is lower than $1$; meaning a more expensive sensor.

\begin{figure}[h]
\small
\begin{tabular}{cc}
\includegraphics[width=0.46\linewidth]{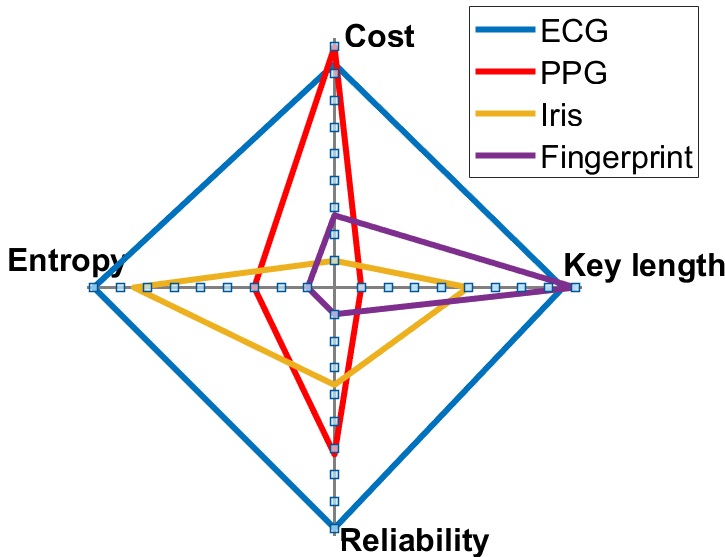}&
\includegraphics[width=0.47\linewidth]{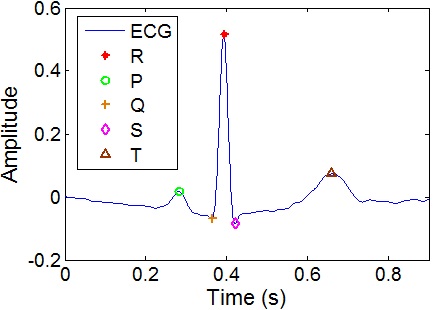}\\
(a) & (b) 
\end{tabular}
\caption{(a) Star graph for comprehensive comparison among four biometrics using NA-IOMBA. (b) One beat ECG with fiducial point.}
\label{cost}
\end{figure}

Figure~\ref{cost} (a) ranks four common technologies (ECG, PPG, iris scan, finger scan) according to four criteria: reliability, entropy, key length, and hardware cost. The maximum point in each length indicates the best candidate for that specific criteria. As can be seen in Figure~\ref{cost}, the average reliability of ECG is $99.76\%$ belong to the maximum point of plot while the average reliability of iris is $98.36\%$ which belongs to the minimum point of plot. In addition, the entropy of ECG is on the maximum point of plot. For the cost, PPG is the best choice among all biometric modalities. Furthermore, the key length of fingerprint is higher than other biometric modalities which made it to become at top of the plot, although the ECG signal is following this with a small margin. ECG appears to be the best candidate for all the criteria when applying NA-IOMBA. However, it is worth noting that ECG still suffers from several other issues (impact of noises, stress condition, and aging) that need to be tackled in order to make this candidate an even stronger selection. NA-IOMBA will be used to alleviate these concerns in the next subsections.

\subsection{Case Study 2: Denoising Overhead Reduction}
In order to determine the sensitivity of ECG key generation based on these feature extraction, the noisy ECG signal with different variances (SNR) has been applied. To view the impact of each noise source, synthetic ECGs are generated and not pre-processed to remove the noise.

\subsubsection{\textbf{ECG synthetic model}} \label{dynamic} 
We also adopt the non-linear dynamical model proposed by McSharry et al.~\cite{mcsharry2003dynamical} to extract dynamic parameters from a real ECG database and generate synthetic ECGs. The aim of this approach is to provide a standard realistic ECG signal with known characteristics from ECG (Fig.~\ref{cost} (b)), which can be generated with specific statistics thereby facilitating the performance evaluation of a given technique. 

\subsubsection{\textbf{ECG Noise Modeling}} \label{sec:NSRSN}
The three main types of noise sources in raw ECG signals are (1) muscle artifacts (MA) which occur due to electrical activity of muscles; (2) baseline wander (BW) caused by body movement; and (3) electrode movement (EM) due to poor contact to the sensor. We adopt the noise model from~\cite{karimian2017noise}.

\begin{figure*}[t]
\begin{tabular}{cc}
\includegraphics[width=0.46\linewidth]{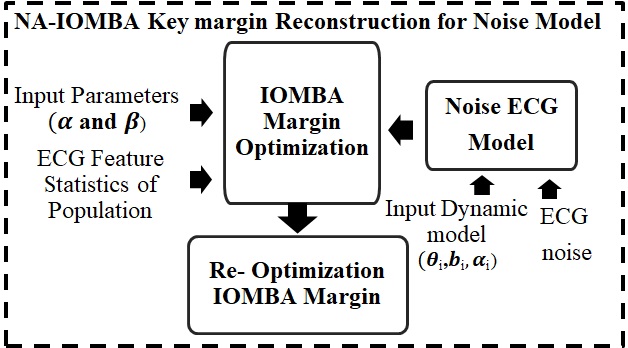}&
\includegraphics[width=0.43\linewidth]{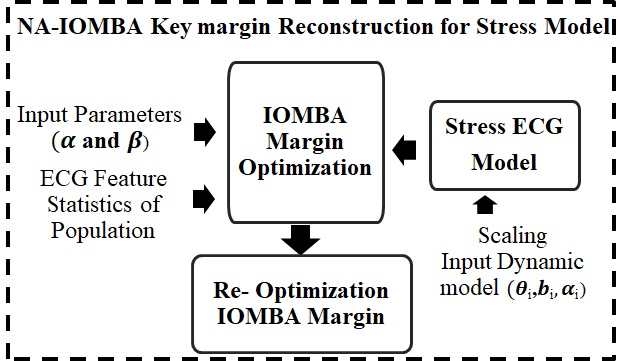}\\
(a) & (b) 
\end{tabular}
\caption{Block diagrams show for NA-IOMBA margin reconstruction key calculation from; (a) noisy ECG , (b) stressed ECG.}
\label{digram}
\end{figure*}

Fig.~\ref{digram} (a) presents the experimental protocol of NA-IOMBA scheme for noisy ECG signals. In noisy ECG key margin reconstruction, dynamic model parameters ($\theta_{i}, \alpha_{i}, b_{i}$) from original ECG signal are considered as the input of the synthetic ECG module. Then, synthetic ECG noise model with desired SNR is employed to add into the clean (synthetic) ECG. 

In order to assess a noise model in NA-IOMBA, we assume $N_e$ and $N_v$  are the noise in the enrollment and verification measurement, respectively. The synthetic noise model is employed as verification noise. It is assumed that $N_e$ and $N_v$ are mutually independent where the standard deviation of measurement noise are denoted as $\sigma_e$ and $\sigma_v$ respectively. We will adopt synthetic ECG noise as our noise model where we assume that $\sigma_v>\sigma_e$. Thus, re-optimization of IOMBA margins module in NA-IOMBA determines new margins based on feedback from this assessment. For our approach, we consider the mixed noises with SNR=5dB for ECG margin reconstruction.

\begin{figure}[h]
\begin{tabular}{cccc}
\includegraphics[width=0.44\linewidth]{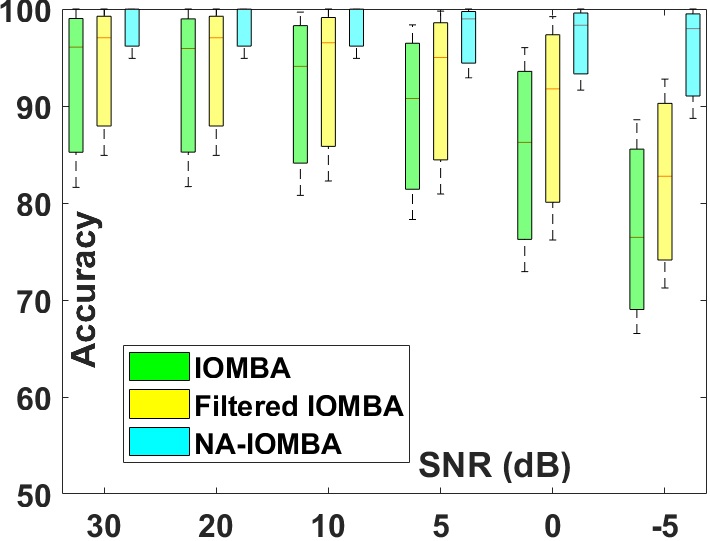}&
\includegraphics[width=0.44\linewidth]{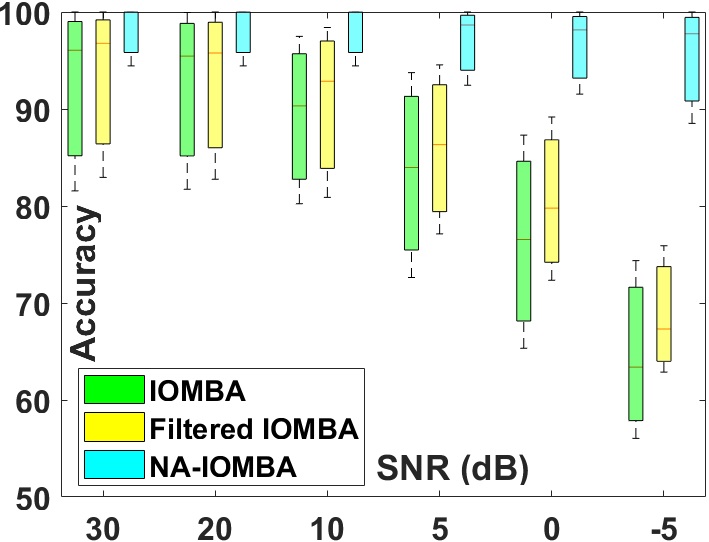}\\
 (c) & (b)\\
\includegraphics[width=0.44\linewidth]{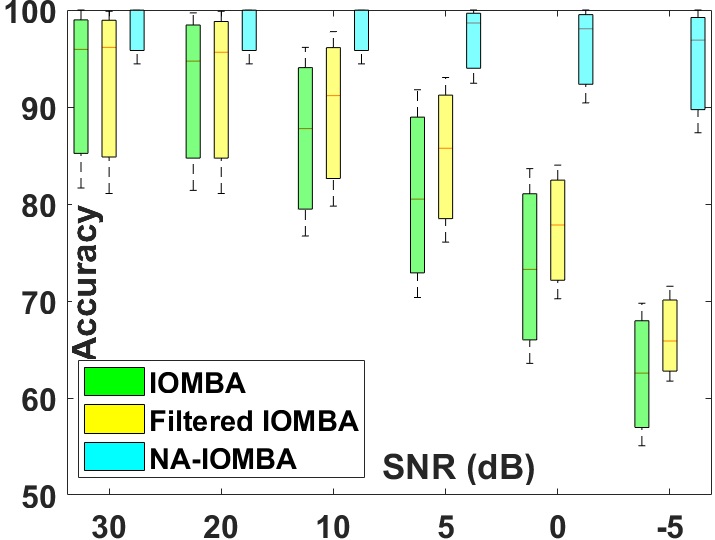}&
\includegraphics[width=0.44\linewidth]{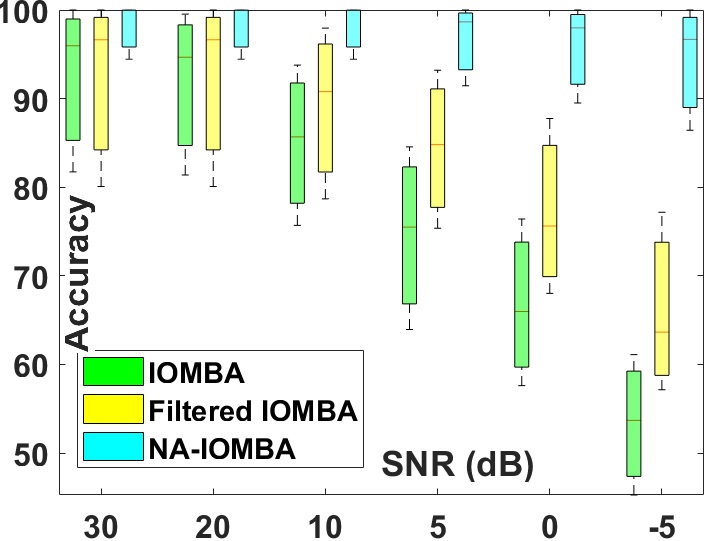}\\
 (c) & (d)
\end{tabular}
\caption{IOMBA, filtered IOMBA, and NA-IOMBA Keys reliability rate vs input SNR; impact of (a) BW noise, (b) EM noise, (c) MA noise, and (d) mixed noises, on the reliability.}
\label{reliability}
\end{figure}

\begin{table*}[t!]
\centering
\caption{Hardware utilization report on the Zynq-7000 SoC XC7Z020.}
\label{fpgal}
%\scalebox{0.7}{
\begin{tabular}{cccc}
\hline
Resource & IOMBA Usage & NA-IOMBA Usage & Available \\ \hline \hline
Flip Flops & 11830 & 1064 & 106400 \\ \hline
Look-Up Tables & 10386 & 2232 & 53200 \\ \hline
DSP Slices & 157 & 21 & 220 \\ \hline
I/O & 32 & 20 & 200 \\ \hline \hline
\end{tabular}
%}
\label{fpga}
\end{table*}

Figure~\ref{reliability} is a box plot showing the reliability rate versus input signal-to-noise ratio (SNR) for ECG based on IOMBA without denoising filters, IOMBA with denoising filters, and NA-IOMBA. The SNR during the noisy segments was set to 30dB, 20dB, 10dB, 5dB, 0dB, and -5dB separately. Figures~\ref{reliability} (a-d) indicate the impact of each noise source (BW, EM, and MA, and mixed of them) on the reliability. In the context of ECG noise levels, the lower SNR provides more fluctuation on ECGs (higher intra-class variation). Intuitively, there appears to be an inverse relationship between the level of generated noise and reliability. Among all these noise sources, MA and EM are the strongest noise sources and have enormous impact on key reliability when IOMBA is applied with and without filtering. As one would expect, IOMBA with filtering obtains better key reliability than without. However, for NA-IOMBA, the reliability is never less than 96.7\% even at worst case (mixed noise with -5dB). In contrast, there is considerable degradation beyond 20dB by using IOMBA with/without filtering (63\% reliability). As mentioned earlier, ECC increases nonlinearly with number of errors. NA-IOMBA has a very high reliability compared to IOMBA, and therefore ECC will inherently consume less overhead. The cost reduction is discussed further below.
\subsubsection{FPGA Implementation of an IOMBA}
In this paper, finite impulse response (FIR) is designed using Simulink in Xilinx System Generator. The Xilinx System Generator tool is a high-level tool for designing high-performance DSP systems and enables us to integrate Xilinx with Simulink. To implement noise reduction using FIR filter, a FDA tool has been applied to design a filter for required specifications. Pan Tompkins algorithm is applied for detecting ECG R peak and segmentation. Finally, the NCN feature extraction technique has been considered for key generation. Table~\ref{fpga} shows implementation of the ECG key generation using IOMBA with filtering and NA-IOMBA without filtering on the Xilinx Zynq-7000. In IOMBA case, 11\% of total flip-flops (FF), 20\% of all available Look-up tables (LUTs), and 71\% of the DSP slice are used while in NA-IOMBA consumes only 1\% of FFs, 4\% of LUTs, and 10\% of DSP are utilized. In addition, IOMBA consumes 113 mW power while NA-IOMBA consumes only 39 mW power. As a result, by saving overall overhead while applying NA-IOMBA, we are able to add ECC in the IoT devices to reconstruct the errors. In fact, NA-IOMBA allows hardware to adapt pre-processing, feature extraction, post-processing, and error correction overheads on a user-to-user basis.

\begin{figure*}[h]
\small
\centering
\begin{tabular}{ccc}
\includegraphics[width=0.3\linewidth]{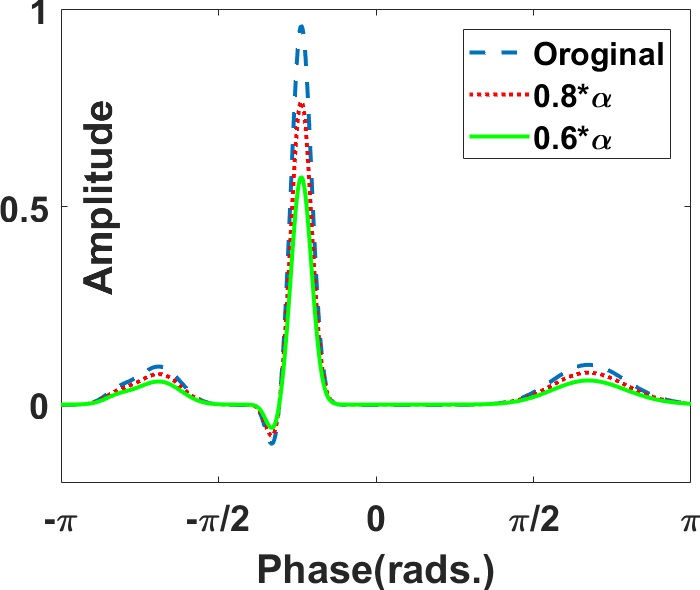} &
\includegraphics[width=0.3\linewidth]{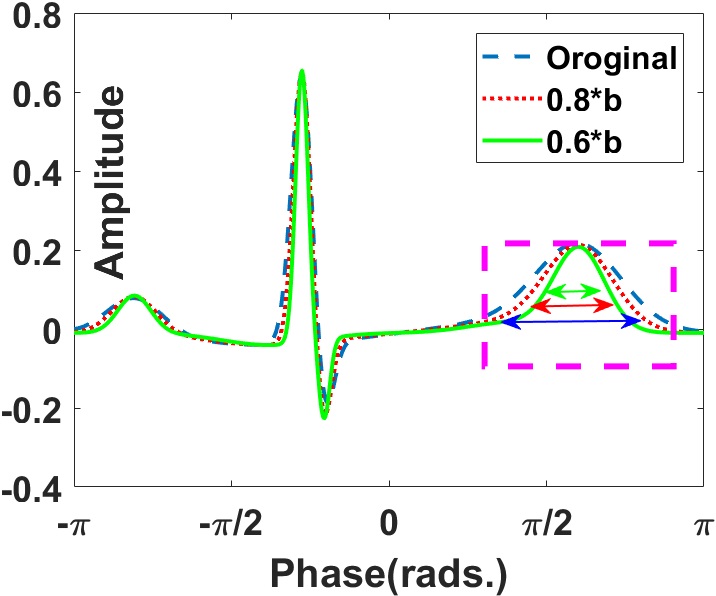} &
\includegraphics[width=0.3\linewidth]{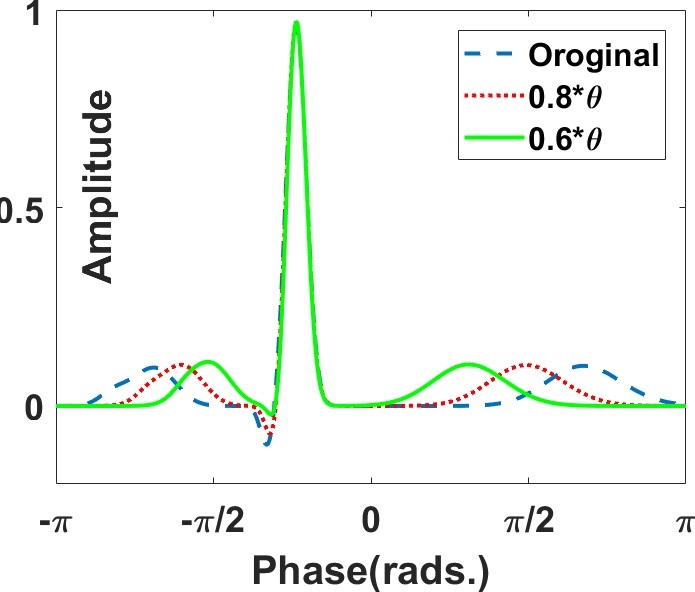} \\
(a) & (b) & (c) 
\end{tabular}
\caption{Impact of ECG signal by decreasing dynamical model parameters: (a) $\alpha$ parameters associated with amplitude, (b) $b$ parameters, and (c) $\theta$ parameters associated with interval and heart rate.}
\label{dynamic}
\end{figure*}

\begin{figure*}[th]
\small
\centering
\begin{tabular}{ccc ccc}
\includegraphics[width=0.3\linewidth]{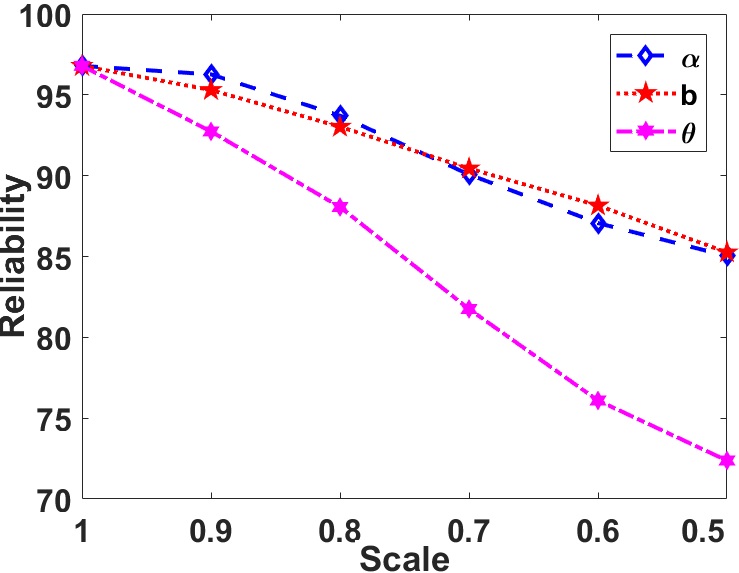} &
\includegraphics[width=0.3\linewidth]{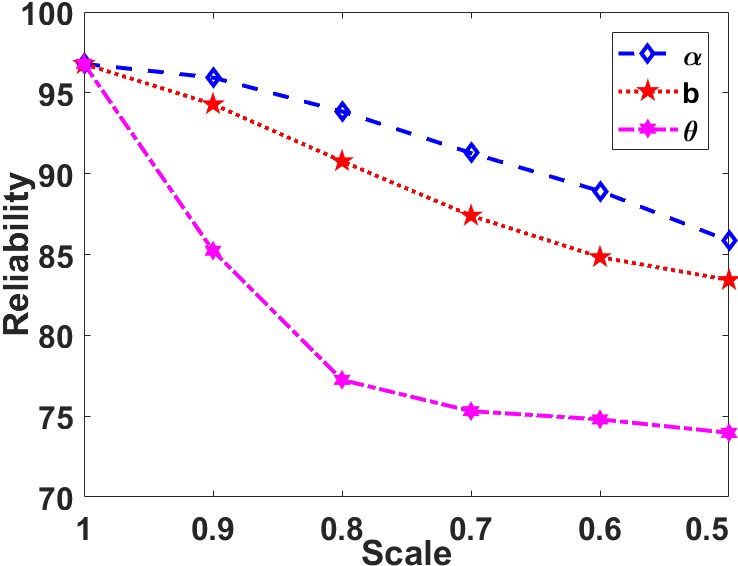} &
\includegraphics[width=0.3\linewidth]{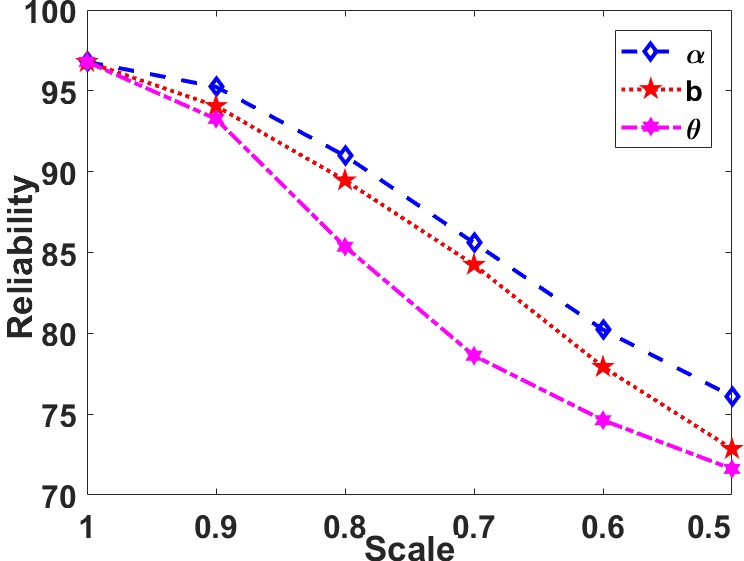}\\
(a) & (b)  & (c) \\
\includegraphics[width=0.3\linewidth]{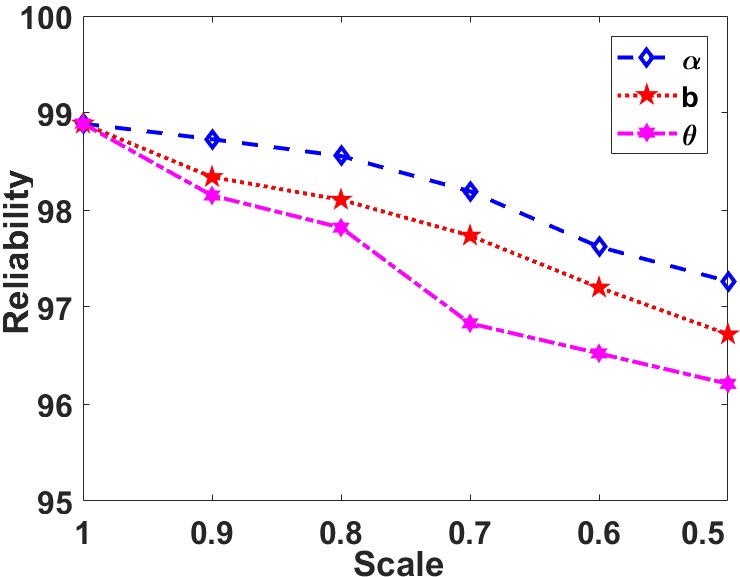} &
\includegraphics[width=0.3\linewidth]{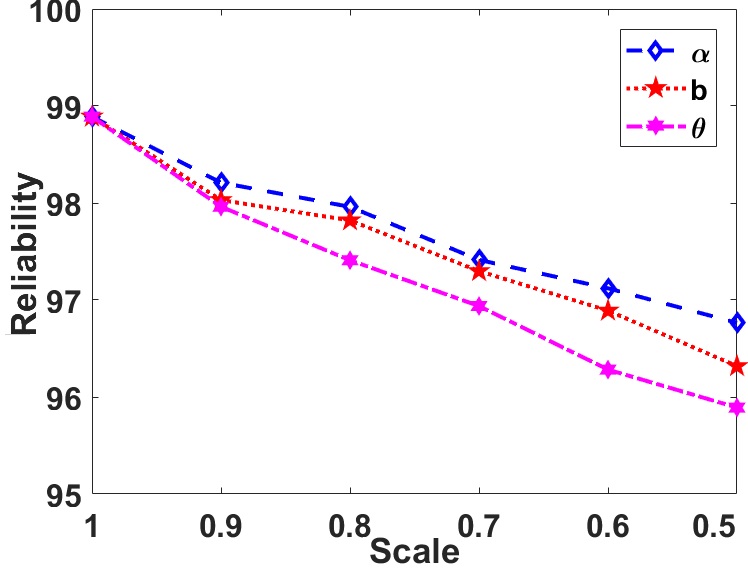} &
\includegraphics[width=0.3\linewidth]{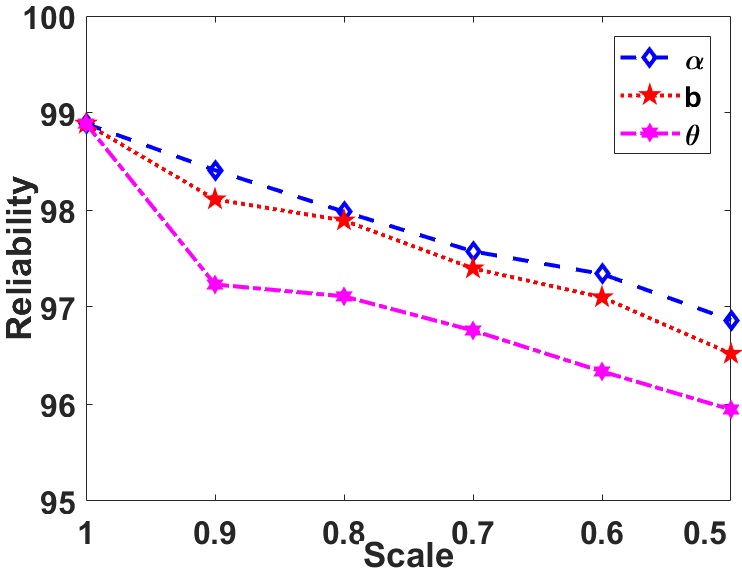}\\
(d) & (e) & (f)
\end{tabular}
\caption{Impact of stress/exercise on the reliability of; IOMBA for changing (a) P wave, (b) QRS wave, (c) T wave; NA-IOMBA for changing, (d) P wave, (e) QRS wave, and (f) T wave of ECG.}
\label{stressmodel}
\end{figure*}

\subsection{Case Study 3: Reducing Enrollment Times}
Another concern that restricts the use of ECG for biometric authentication is the variability of cardiac rhythm within the subjects. Heart rate varies with individual’s physiological and mental conditions. Stress, excitement, exercise, and other working activities may have impact on the heart rate and can elevate it. These variations are likely to affect the reliability of ECG based key generation. A previous study~\cite{kim2006robust} about the influences of physical exercises indicates that the ECG morphology is affected by exercise/stress. In other words, each peak (P, QRS, T) in the ECG may increase/decrease in amplitude, temporal location, etc. To cover the impact of stress/exercise on the reliability with different scenario, we vary the dynamical model parameters by type ($\theta_{i}, \alpha_{i}, b_{i} \forall i$) and analyze the impact of each part of ECG waveform. In Fig.~\ref{digram} (b), we illustrate how noise in ECG features from stress/exercise can be handled using the model in NA-IOMBA. In order to assess the impact of stress in NA-IOMBA, NA-IOMBA trains itself with the help of the information from standard deviation of stress ECG signal model and re-optimizes IOMBA margins.

Each dynamical model parameter is scaled by a factor ($0.9-0.5$). Fig.~\ref{dynamic}(a-c) show how the ECG changes when scaling $\alpha$, $b$, and $\theta$ parameters respectively. As shown in Fig.~\ref{dynamic} (a), the $\alpha$ parameter controls amplitudes of each component of ECG waveform. In contrast, the onset and offset of ECG waveform and interval duration are associated with scaling $b$ and $\theta$ parameters. We intend to investigate the impact of dynamic parameters on each ECG waveform component to analyze the impact of stress \& exercise on the reliability of IOMBA/NA-IOMBA. 

Fig.~\ref{stressmodel}((a-c) indicate the reliability of each dynamic parameters on the P, QRS, and T waves based on IOMBA, respectively. The reliability of NA-IOMBA is shown in Fig.~\ref{stressmodel}(d-f) where IOMBA margins are re-optimized assuming dynamic parameters scaled to $0.7$. For IOMBA, the T wave is impacted by all parameters ($\alpha$, $b$, and $\theta$). At lowest scale value (0.5), the minimum reliability of dynamic parameters of T wave is $71.61\%$ while for P and QRS wave are $72.37\%$, and $73.96\%$ respectively. $\theta$ has a larger impact than other parameters for P and QRS waves because it causes distortion in ECG time intervals (distances between peaks). Even though there is too much degradation on the reliability for IOMBA technique, the performance of NA-IOMBA in stress/exercise situation is much higher than IOMBA. The minimum reliability of dynamic parameter on P, QRS, and T waves (Fig.~\ref{stressmodel}(d)) are improved by 25\%. Note, however, that the key length is sacrificed by 56\% (420 average key bits) to obtain this improvement.  We expect that this will be long enough for most cryptographic applications as well as BLOcKeR.

\section{{Conclusions}}
In this paper, we introduced the first ever biometric system implementation that utilizes reconfigurability, bitstream obfuscation, strong PUFs, and PUF challenge-response predication models. The non-invertibility, non-linkability, and information security of keys used by the proposed approach made it resistant to many common attacks against match-on card/device biometric systems. In addition, the interval optimized mapping bit allocation (IOMBA) scheme for key generation was improved by incorporating noise models. It was demonstrated that keys generated from ECG, PPG, iris, and fingerprint by noise-aware IOMBA were more reliable, longer, and higher entropy than noise-free IOMBA. In addition, by using more advanced noise models for ECG, overhead from denoising filters and error correction could be further reduced by 62\%  without additional enrollment measurements. In future, we will develop noise-tolerant bitstream obfuscation approaches and analyze them with respect to overhead and vulnerability to attacks.


\begin{thebibliography}{10}

\bibitem{sameni2005filtering}
R.~Sameni, M.~B. Shamsollahi, C.~Jutten, M.~Babaie-Zade,
``Filtering noisy ECG signals using the extended Kalman filter based on a modified dynamic ECG model,'' in \emph{Computers in Cardiology 2005}.\hskip 1em plus 0.5em minus 0.4em\relax IEEE, 2005, pp. 1017--1020.

\bibitem{karimian2017noise}
N.~Karimian, F.~Tehranipoor, Z.~Guo, M.~M. Tehranipoor, D.~Forte,
``Noise assessment framework for optimizing ECG key generation,'' in \emph{Technologies for Homeland Security (HST), 2017 IEEE International Symposium on}.\hskip 1em plus 0.5em minus 0.4em\relax IEEE, 2017, pp. 1--6

\bibitem{karimian2017highly}
N.~Karimian, Z.~Guo, M.~M. Tehranipoor, D.~Forte,
``Highly reliable key generation from electrocardiogram (ECG),'' \emph{IEEE Transactions on Biomedical Engineering}, vol.~64, no.~6, pp. 1400--1411, 2017.

\bibitem{wubbeler2007verification}
G.~W{\"u}bbeler, M.~Stavridis, D.~Kreiseler, R.~D. Bousseljot, C.~Elster,
``Verification of humans using the electrocardiogram,'' \emph{Pattern Recognition Letters}, vol.~28, no.~10, pp. 1172--1175, 2007.


\bibitem{jain201650}
A.~Jain, K.~Nandakumar, A.~Ross,
``50 years of biometric research: Accomplishments, challenges, and opportunities,'' \emph{Pattern Recognition Letters}, vol.~79, pp. 80--105, 2016.


\bibitem{fernandes2017novel}
S.~L. Fernandes, V.~P. Nandakumar, N.~R. Sunder, N.~Arunkumar, S.~Kadry,
``A novel nonintrusive decision support approach for heart rate measurement,'' \emph{Pattern Recognition Letters}, 2017.


\bibitem{hong1998fingerprint}
S.~L. Fernandes, V.~P. Nandakumar, N.~R. Sunder, N.~Arunkumar, S.~Kadry,
``Fingerprint image enhancement: Algorithm and performance evaluation,'' \emph{IEEE transactions on pattern analysis and machine intelligence}, vol.~20, no.~8, pp. 777--789, 1998.

\bibitem{bowyer2016handbook}
K.~W. Bowyer, M.~J. Burge, \emph{Handbook of iris recognition}.\hskip 1em plus 0.5em minus 0.4em\relax Springer, 2016.


\bibitem{ross2003hybrid}
A.~Ross, A.~Jain, J.~Reisman,
``A hybrid fingerprint matcher,'' \emph{Pattern Recognition}, vol.~36, no.~7, pp. 1661--1673, 2003.

\bibitem{maio2004fvc2004}
D.~Maio, D.~Maltoni, R.~Cappelli, J.~Wayman, A.~Jain,
``FVC2004: Third fingerprint verification competition,'' \emph{Springer}, pp. 31--35, 2004.

\bibitem{maltoni2009handbook}
 D.~Maltoni, D.~Maio, A.~Jain, S.~Prabhakar, \emph{Handbook of fingerprint recognition}.\hskip 1em plus 0.5em minus 0.4em\relax Springer Science \& Business Media, 2009.

\bibitem{karlen2013multiparameter}
W.~Karlen, S.~Raman, J.~M. Ansermino, G.~Dumont,
``Multiparameter respiratory rate estimation from the photoplethysmogram,'' \emph{IEEE Transactions on Biomedical Engineering}, vol.~60, no.~7, pp. 1946--1953, 2013.

\bibitem{CASIA}
``{CASIA-IrisV1},'' { http://biometrics.idealtest.org}, accessed March 2018.

\bibitem{goldberger2000physiobank}
``{Physiobank},'' {http://physionet.org}, accessed March 2018.

\bibitem{rose2016picking}
``{Picking bluetooth low energy locks from a quarter mile away},'' {https://media.defcon.org/DEF\%20CON\%2024/DEF\%
20CON\%2024\%20presentations/DEFCON-24-RoseRamsey-Picking-Bluetooth-Low-Energy-LocksUPDATED.pdf.}, accessed March 2018.

\bibitem{Backdooring}
``{Backdooring the Frontdoor},'' {https://media.defcon.org}, accessed March 2018.

\bibitem{agrafioti2012secure}
F.~Agrafioti, B.~Francis, D.~Hatzinakos,
``Secure telemedicine: Biometrics for remote and continuous patient verification,'' \emph{Journal of Computer Networks and Communications}, 2012.

\bibitem{jain2008biometric}
A.~Jain, K.~Nandakumar, N.~Abhishek,
``Biometric template security,'' \emph{EURASIP Journal on Advances in Signal Processing}, vol.~8, 2008.


\bibitem{adler2005vulnerabilities}
A.~Adler,
``Vulnerabilities in biometric encryption systems,'' in \emph{International Conference on Audio-and Video-Based Biometric Person Authentication}.\hskip 1em plus 0.5em minus 0.4em\relax IEEE, 2005, pp. 1100--1109.


\bibitem{Equifax}
``{At Equifax, a Category 5 data breach},'' {www.usatoday.com}, accessed March 2018.

\bibitem{washingtonpost}
``{5.6 million fingerprints stolen in cyberattack},'' {https://www.washingtonpost.com}, accessed March 2018.

\bibitem{techcrunch}
``{5.6 million fingerprints stolen in cyberattack},'' {https://www.washingtonpost.com}, accessed March 2018.

\bibitem{cheon2016ghostshell}
J.~Cheon, H.~Chung, M.~ Kim, K.~lee,
``Ghostshell: Secure Biometric Authentication using Integrity-based Homomorphic Evaluations,'' \emph{IACR Cryptology ePrint Archive}, 2016.


\bibitem{karabat2015thrive}
A.~Adler,
``THRIVE: threshold homomorphic encryption based secure and privacy preserving biometric verification system,'' \emph{EURASIP Journal on Advances in Signal Processing}, no.~1, pp. 1--71, 2015.


\bibitem{Cio}
``{How to make Fully Homomorphic Encryption `practical and usable'},'' {https://www.cio.com/article/3196872/security}, accessed March 2018.

\bibitem{forte2017hardware}
D.~Forte, S.~Bhunia, and M.~M. Tehranipoor, \emph{Hardware protection through
  obfuscation}.\hskip 1em plus 0.5em minus 0.4em\relax Springer, 2017.

\bibitem{gassend2002silicon}
B.~Gassend, D.~Clarke, M.~Dijk, S.~Devadas,
``Silicon physical random functions,'' in \emph{Proceedings of the 9th ACM conference on Computer and communications security}.\hskip 1em plus 0.5em minus 0.4em\relax ACM, 2002, pp. 148--160.

\bibitem{suh2007physical}
G.~Suh, S.~Devadas,
``Physical unclonable functions for device authentication and secret key generation,'' in \emph{Proceedings of the 44th annual design automation conference}.\hskip 1em plus 0.5em minus 0.4em\relax ACM, 2007, pp. 9--14.


\bibitem{chakraborty2009harpoon}
S.~L. Fernandes, V.~P. Nandakumar, N.~R. Sunder, N.~Arunkumar, S.~Kadry,
``HARPOON: an obfuscation-based SoC design methodology for hardware protection,'' \emph{IEEE Transactions on Computer-Aided Design of Integrated Circuits and Systems}, vol.~28, no.~10, pp. 1493--1502, 2009.

\bibitem{guo2015investigation}
Z.~Guo, M.~Tehranipoor, D.~Forte, J.~Di,
``Investigation of obfuscation-based anti-reverse engineering for printed circuit boards,'' in \emph{Proceedings of the 52nd Annual Design Automation Conference}.\hskip 1em plus 0.5em minus 0.4em\relax ACM, 2015.


\bibitem{karam2016robust}
R.~Karam, T.~Hoque, S.~Ray, M.~Tehranipoor, S.~Bhunia,
``Investigation of obfuscation-based anti-reverse engineering for printed circuit boards,'' in \emph{ReConFigurable Computing and FPGAs (ReConFig), 2016 International Conference on}.\hskip 1em plus 0.5em minus 0.4em\relax IEEE, pp. 1--8, 2016.


\bibitem{maes2016physically}
R.~Maes, \emph{Physically Unclonable Functions}.\hskip 1em plus 0.5em minus 0.4em\relax Springer, 2016.



\bibitem{marinissen2014direct}
E.~Marinissen, B.~Wachter, K.~Smith, J.~Kiesewetter, M.~Taouil, S.~Hamdioui,
``Direct probing on large-array fine-pitch micro-bumps of a wide-I/O logic-memory interface,'' in \emph{Test Conference (ITC), 2014 IEEE International}.\hskip 1em plus 0.5em minus 0.4em\relax IEEE, pp. 1--10, 2014.


\bibitem{rajendran2012logic}
J.~Rajendran, Y.~Pino, O.~Sinanoglu, R.~Karri,
``Direct probing on large-array fine-pitch micro-bumps of a wide-I/O logic-memory interface,'' in \emph{Proceedings of the Conference on Design, Automation and Test in Europe}.\hskip 1em plus 0.5em minus 0.4em\relax EDA Consortium, pp. 953--958, 2012.


\bibitem{ruhrmair2010modeling}
U.~Ruhrmair, F.~Sehnke, G.~Dror, S.~Devadas, J.~Schmidhuber,
``Modeling attacks on physical unclonable functions,'' in \emph{Proceedings of the 17th ACM conference on Computer and communications security}.\hskip 1em plus 0.5em minus 0.4em\relax ACM, pp. 237--249, 2010.


\bibitem{skorobogatov2010flash}
S.~Skorobogatov,
``Flash memory ‘bumping’attacks,'' in \emph{Cryptographic Hardware and Embedded Systems, CHES 2010}.\hskip 1em plus 0.5em minus 0.4em\relax Springer, pp. 158--172, 2010.

\bibitem{skorobogatov2005semi}
S.~P Skorobogatov, \emph{Semi-invasive attacks: a new approach to hardware security analysis}.\hskip 1em plus 0.5em minus 0.4em\relax University of Cambridge Ph. D. dissertation, 2005.


\bibitem{weber2005software}
S.~Weber, P.~Karger, A.~Paradkar
``A software flaw taxonomy: aiming tools at security,'' in \emph{ACM SIGSOFT Software Engineering Notes}.\hskip 1em plus 0.5em minus 0.4em\relax ACM, vol.~30, no.~4, pp. 1--7, 2005.

\bibitem{yang2004scan}
B.~Yang, K.~Wu, R.~Karri
``Scan based side channel attack on dedicated hardware implementations of data encryption standard,'' in \emph{Test Conference, 2004. Proceedings. ITC 2004. International}.\hskip 1em plus 0.5em minus 0.4em\relax IEEE, pp. 339--344, 2004.


\bibitem{stewart2016biometrics}
``{Biometrics and Privacy: On Device vs On Server matching},'' {https://go.noknok.com/rs/207-VEO-726/images/PwCLegal-Biometric-Privacy.pdf}, accessed March 2018.


\bibitem{mcsharry2003dynamical}
P.~McSharry, G.~Clifford, L.~Tarassenko, L.~Smith,
``A dynamical model for generating synthetic electrocardiogram signals,'' \emph{IEEE transactions on biomedical engineering}, vol.~50, no.~3, pp. 289--294, 2003.


\bibitem{sameni2007multichannel}
R.~Sameni, G.~Clifford, C.~Jutten, M.~Shamsollahi,
``Multichannel ECG and noise modeling: Application to maternal and fetal ECG signals,'' \emph{EURASIP Journal on Applied Signal Processing}, no.~1, pp. 94--94, 2007.


\bibitem{karimian2017human}
N.~Karimian, M.~Tehranipoor, D.~Forte, 
``Non-fiducial ppg-based authentication for healthcare application,'' in \emph{Biomedical \& Health Informatics (BHI), 2017 IEEE EMBS International Conference on}.\hskip 1em plus 0.5em minus 0.4em\relax IEEE, pp. 429--4324, 2017.

\bibitem{daugman2004iris}
John.~Daugman,
``How iris recognition works,'' \emph{IEEE Transactions on circuits and systems for video technology}, vol.~14, no.~1, pp. 21--30, 2004.


\bibitem{subramanyan2015evaluating}
P.~Subramanyan, S.~Ray, S.~Malik, 
``Evaluating the security of logic encryption algorithms,'' in \emph{Hardware Oriented Security and Trust (HOST), 2015 IEEE International Symposium on}.\hskip 1em plus 0.5em minus 0.4em\relax IEEE, pp. 137--143, 2015.

\bibitem{kelsey1998side}
J.~Kelsey, B.~Schneier, W.~Wagner, C.~Hall,
``Side channel cryptanalysis of product ciphers,'' in \emph{European Symposium on Research in Computer Security}.\hskip 1em plus 0.5em minus 0.4em\relax Springer, pp. 97--110, 1998.

\bibitem{maiorana2015hill}
M.~Maiorana, G.~Hine, P.~Campisi,
``Hill-climbing attacks on multibiometrics recognition systems,'' \emph{IEEE Transactions on Information Forensics and Security}, vol.~10, no.~10, pp. 900--915, 2015.

\bibitem{kim2006robust}
K.~Kim, T.~Yoon, J.~Lee, D.~Kim, H.~Koo
``A robust human identification by normalized time-domain features of electrocardiogram,'' in \emph{Engineering in medicine and biology society, .27th annual international conference of the}.\hskip 1em plus 0.5em minus 0.4em\relax IEEE, pp. 1114--1117, 2006.


%\bibitem{11-Book-obfuscation}
%D.~Forte, S.~Bhunia, and M.~M. Tehranipoor, \emph{Hardware protection through
%  obfuscation}.\hskip 1em plus 0.5em minus 0.4em\relax Springer, 2017.
%
%\bibitem{5-Ankur-DAC}
%Y.~Xie and A.~Srivastava, ``Delay locking: Security enhancement of logic
%  locking against ic counterfeiting and overproduction,'' in \emph{Proc. Design Automation Conference 2017}.\hskip 1em plus 0.5em
%  minus 0.4em\relax ACM, 2017, p.~9.
%
%\bibitem{4-Malik-HOST}
%P.~Subramanyan, S.~Ray, and S.~Malik, ``Evaluating the security of logic
%  encryption algorithms,'' in \emph{Proc. Hardware Oriented Security and Trust
%  (HOST)}.\hskip 1em plus 0.5em minus
%  0.4em\relax IEEE, 2015, pp. 137--143.
%
%\bibitem{10-Rajjendran-DAC}
%J.~Rajendran, Y.~Pino, O.~Sinanoglu, and R.~Karri, ``Logic encryption: A fault
%  analysis perspective,'' in \emph{Proc. Conf. Design,
%  Automation and Test in Europe}.\hskip 1em plus 0.5em minus 0.4em\relax EDA
%  Consortium, 2012, pp. 953--958.
%
%\bibitem{12-Plaza-TCAD}
%S.~M. Plaza and I.~L. Markov, ``Solving the third-shift problem in ic piracy
%  with test-aware logic locking,'' \emph{IEEE Trans. Computer-Aided
%  Design of Integrated Circuits and Systems}, vol.~34, no.~6, pp. 961--971,
%  2015.
%
%\bibitem{last1}
%K.~Shamsi, M.~Li, T.~Meade, Z.~Zhao, D.~Z. Pan, and Y.~Jin, ``Cyclic
%  obfuscation for creating sat-unresolvable circuits,'' in \emph{Proc.
%  the on Great Lakes Symposium on VLSI}.\hskip 1em plus 0.5em minus
%  0.4em\relax ACM, 2017, pp. 173--178.
%
%\bibitem{last2}
%M.~Yasin, B.~Mazumdar, O.~Sinanoglu, and J.~Rajendran, ``Removal attacks on
%  logic locking and camouflaging techniques,'' \emph{IEEE Trans.
%  Emerging Topics in Computing}, 2018, to appear.
%
%\bibitem{last3}
%X.~Xu, B.~Shakya, M.~M. Tehranipoor, and D.~Forte, ``Novel bypass attack and
%  bdd-based tradeoff analysis against all known logic locking attacks,'' in
%  \emph{Proc. Int. Conf. Cryptographic Hardware and Embedded
%  Systems}.\hskip 1em plus 0.5em minus 0.4em\relax Springer, 2017, pp.
%  189--210.
%
%\bibitem{Goodfellow}
%I.~Goodfellow, Y.~Bengio, A.~Courville, and Y.~Bengio, \emph{Deep
%  learning}.\hskip 1em plus 0.5em minus 0.4em\relax MIT press Cambridge, 2016,
%  vol.~1.
%
%\bibitem{1-bhunia-DAC}
%R.~S. Chakraborty and S.~Bhunia, ``Hardware protection and authentication
%  through netlist level obfuscation,'' in \emph{Proc. IEEE/ACM International Conference on Computer-Aided Design}.\hskip 1em plus
%  0.5em minus 0.4em\relax IEEE Press, 2008, pp. 674--677.
%
%\bibitem{2-karri-DAC}
%J.~Rajendran, Y.~Pino, O.~Sinanoglu, and R.~Karri, ``Security analysis of logic
%  obfuscation,'' in \emph{Proc. Design Automation
%  Conference}.\hskip 1em plus 0.5em minus 0.4em\relax ACM, 2012, pp. 83--89.
%
%\bibitem{3-farinaz-DATE}
%J.~A. Roy, F.~Koushanfar, and L.~Igor, ``Ending piracy of
%  integrated circuits,'' in \emph{Proc. Conf. Design,
%  Automation, and Test in Europe}, 2008, pp. 1069--1074.
%
%\bibitem{6-Ankur-CHES}
%Y.~Xie and A.~Srivastava, ``Mitigating sat attack on logic locking,'' in
%  \emph{Proc. Int. Conf. Cryptographic Hardware and Embedded
%  Systems}.\hskip 1em plus 0.5em minus 0.4em\relax Springer, 2016, pp.
%  127--146.
%
%\bibitem{7-Yasin-HOST}
%M.~Yasin, B.~Mazumdar, J.~J. Rajendran, and O.~Sinanoglu, ``Sarlock: Sat attack
%  resistant logic locking,'' in \emph{Proc. Hardware Oriented Security and Trust
%  (HOST)}.\hskip 1em plus 0.5em minus
%  0.4em\relax IEEE, 2016, pp. 236--241.
%
%\bibitem{shamsi2017appsat}
%K.~Shamsi, M.~Li, T.~Meade, Z.~Zhao, D.~Z. Pan, and Y.~Jin, ``Appsat:
%  Approximately deobfuscating integrated circuits,'' in \emph{Proc. Hardware Oriented
%  Security and Trust (HOST)}.\hskip 1em
%  plus 0.5em minus 0.4em\relax IEEE, 2017, pp. 95--100.
%
%\bibitem{8-benchmark}
%``{Trust-Hub},'' {https:/trust-hub.org/OBFbenchmarks.html}, accessed March 2018.
%
%\bibitem{9-systemverilog}
%C.~Spear, ``A guide to learning the testbench language features,'' \emph{System
%  Verilog for verification, 2nd ed., Springer Publishing Company,
%  Incorporated}, pp. 11--18, 2008.
%
%\bibitem{Alex}
%A.~Graves, A.-r. Mohamed, and G.~Hinton, ``Speech recognition with deep
%  recurrent neural networks,'' in \emph{Proc. Acoustics, Speech and Signal Processing}.\hskip 1em plus 0.5em minus
%  0.4em\relax IEEE, 2013, pp. 6645--6649.
%
%\bibitem{Sutskever}
%I.~Sutskever, O.~Vinyals, and Q.~V. Le, ``Sequence to sequence learning with
%  neural networks,'' in \emph{Proc. Advances in Neural Information Processing
%  Systems}, 2014, pp. 3104--3112.
%
%\bibitem{Hinton}
%Y.~LeCun, Y.~Bengio, and G.~Hinton, ``Deep learning,'' \emph{Nature}, vol. 521,
%  no. 7553, p. 436, 2015.
%
%\bibitem{13-Hochreiter}
%S.~Hochreiter and J.~Schmidhuber, ``Long short-term memory,'' \emph{Neural
%  Computation}, vol.~9, no.~8, pp. 1735--1780, 1997.
\end{thebibliography}
\end{document}